\documentclass[journal=aamick,manuscript=article]{achemso}

\usepackage[version=3]{mhchem} 
\usepackage{braket}
\usepackage{color}

\author{Daniel Duarte-Ruiz}
\affiliation{Carl von Ossietzky Universit\"at Oldenburg, Institute of Physics, 26129 Oldenburg, Germany}
\altaffiliation{These authors contributed equally to this work.}
\author{Timo Reents}
\affiliation{Carl von Ossietzky Universit\"at Oldenburg, Institute of Physics, 26129 Oldenburg, Germany}
\alsoaffiliation{PSI Center for Scientific Computing, Theory and Data, Paul Scherrer Institute, 5232 Villigen PSI, Switzerland}
\altaffiliation{These authors contributed equally to this work.}
\author{Elmar Yu. Kataev}
\affiliation{Department Interface Design, Helmholtz-Zentrum Berlin f\"ur Materialien und Energie GmbH, Berlin, Germany}
\alsoaffiliation{Department of Chemistry and Pharmacy, Friedrich-Alexander-Universität Erlangen-Nürnberg, Erlangen, Germany}
\author{Raul Garcia-Diez}
\affiliation{Department Interface Design, Helmholtz-Zentrum Berlin f\"ur Materialien und Energie GmbH, Berlin, Germany}
\author{Regan G. Wilks}
\affiliation{Department Interface Design, Helmholtz-Zentrum Berlin f\"ur Materialien und Energie GmbH, Berlin, Germany}
\alsoaffiliation{Energy Materials In-Situ Laboratory Berlin (EMIL), Helmholtz-Zentrum Berlin f\"ur Materialien und Energie 
GmbH, Berlin, Germany}
\author{Marcus B\"ar}
\affiliation{Department Interface Design, Helmholtz-Zentrum Berlin f\"ur Materialien und Energie GmbH, Berlin, Germany}
\alsoaffiliation{Energy Materials In-Situ Laboratory Berlin (EMIL), Helmholtz-Zentrum Berlin f\"ur Materialien und Energie 
GmbH, Berlin, Germany}
\alsoaffiliation{Department of Chemistry and Pharmacy, Friedrich-Alexander-Universität Erlangen-Nürnberg, Erlangen, Germany}
\alsoaffiliation{Helmholtz-Institute Erlangen-Nürnberg for Renewable Energy (HI ERN), Berlin, Germany}
\author{Caterina Cocchi}
\affiliation{Friedrich-Schiller-Universit\"at Jena, Institute of Condensed Matter Theory and Optics, 07743 Jena, Germany}
\alsoaffiliation{Friedrich-Schiller-Universit\"at Jena, Abbe Center of Photonics, 07745 Jena, Germany}
\alsoaffiliation{Carl von Ossietzky Universit\"at Oldenburg, Institute of Physics, 26129 Oldenburg, Germany}
\email{caterina.cocchi@uni-jena.de}

\title{Decoding Oxygen K-edge Fingerprints of NCM-811 Degradation via Ab Initio Many-Body Theory and High-Throughput Screening of Crystal Proxies}

\abbreviations{IR,NMR,UV}

\begin{document}

\begin{tocentry}
\includegraphics[width=3.25in]{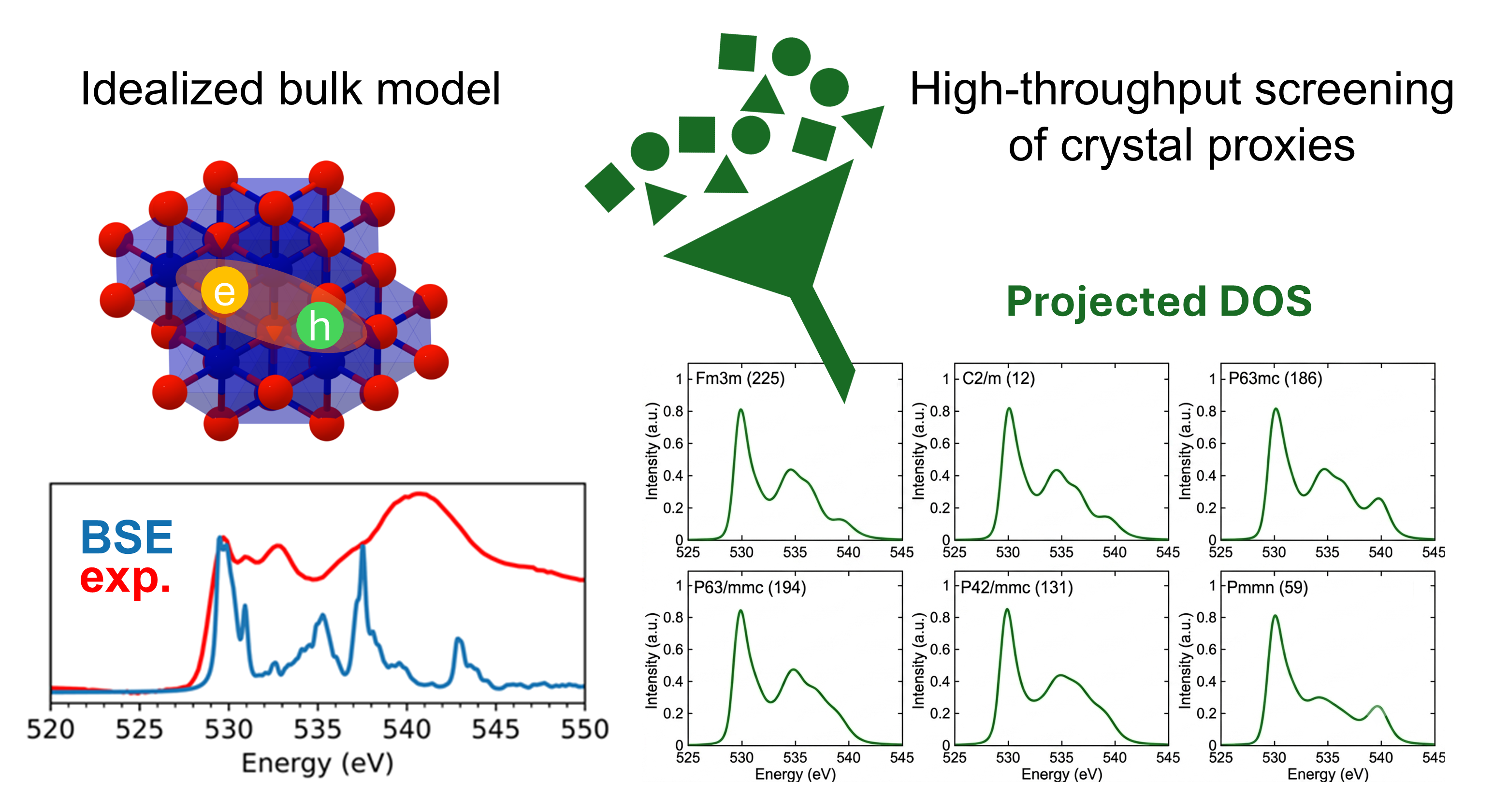}
\end{tocentry}

\newpage
\begin{abstract}

  The degradation of \ce{LiNi_{0.8}Co_{0.1}Mn_{0.1}O2} (NCM-811) in Li-ion batteries produces complex transition-metal oxides and binary phases that fundamentally limit cathode performance. While identifying these degradation products via X-ray absorption spectroscopy (XAS) is essential for mitigating electrochemical performance loss, interpretation remains challenging due to the structural complexity of real-world samples. In this work, we present an integrated theoretical-experimental framework combining high-resolution oxygen K-edge XAS with \textit{ab initio} simulations based on many-body perturbation theory and high-throughput screening from density functional theory. We first evaluate the spectroscopic fingerprints of eight layered, spinel, and nominal rock-salt reference oxides, identifying discrepancies between the idealized single-crystal bulk phase and experimental spectra. Using high-throughput screening to analyze the oxygen $p$-projected density of states of 38 distinct polymorphs of \ce{NiO}, \ce{CoO}, and \ce{MnO}, we propose that the spectral differences can emerge, among other factors, from local structural variations represented here by simplified structural proxies. Our work establishes a viable and rigorous computational pathway to interpret the complex landscape of degraded battery materials.

\end{abstract}

\newpage
\section{Introduction}
Nickel cobalt manganese oxides (NCM) are a cornerstone of modern Li-ion battery technology~\cite{debi+19am,fich+22aem}. The intrinsic structural and chemical complexity of these systems~\cite{debi+17jpcc,jia+22nl} is further enhanced during electrochemical cycling. In Ni-rich compositions such as \ce{LiNi_{0.8}Co_{0.1}Mn_{0.1}O_2} (NCM-811), phase transitions~\cite{kond+17jpcc} modify the layered structure of the cathode material into spinel and rock-salt crystals~\cite{zou+18acsel,hua+19natcom,chen+21nano,gan+23nr}. Moreover, oxygen release during charge at higher cut-off voltages triggers the formation of transition-metal (TM) oxide phases as degradation products~\cite{hu+21acsami}, which lead to loss of reversible Li storage, impede Li$^+$ and electronic transport, and promote TM dissolution and electrolyte decomposition~\cite{huan+21acsami,luec+23acsomega}. 
Gaining an in-depth understanding of these parasitic phases is therefore of paramount importance for elucidating aging pathways and optimizing Ni-rich cathode performance~\cite{jung+14aem,dixi+17jpcc,ryu+18cm,edge+21pccp}.

Experimentally, oxygen loss and TM dissolution drive irreversible reconstruction of Ni-rich and Li-rich layered cathodes, generating spinel-like and rock salt-like regions with altered metal-oxygen coordination~\cite{debi+19am,zuo+20ees,lu+24csr,zhen+25afm,liu+26ees}. The growth of these poorly conducting surface layers impedes Li-ion transport and increases electrode impedance, directly connecting structural reconstruction to electrochemical degradation. Atomic-resolution transmission electron microscopy (TEM) and spatially resolved electron energy loss spectroscopy reveal cation redistribution, oxygen depletion, and changes in TM valence across reconstructed regions, while in situ TEM resolves the progression of structural transformations~\cite{yin+22afm,xia+23jmst}. In Ni-rich cathodes, anisotropic lattice contraction additionally generates mechanical stress and microcracking, exposing fresh surfaces to electrolyte reactions and promoting bulk instability through interfacial degradation~\cite{debi+19am}. Together, these observations establish oxygen deficiency, cation disorder, and local coordination as structural variables for interpreting spectroscopic signatures of degradation products. This provides a physical basis for utilizing reference binary and ternary oxides as benchmarks and local structural proxies employed here.

X-ray absorption spectroscopy (XAS) is a powerful tool for characterizing complex cathode materials both \textit{in situ} and \textit{ex situ}~\cite{john+02ec}. Thanks to its unique sensitivity to local atomic environments and electronic symmetry without the requirement of long-range order~\cite{cocc+16prb,hua+19natcom}, this technique can resolve the fingerprints of coexisting and/or disordered phases. In this context, probing the oxygen $K$-edge is particularly informative, as its low-energy features arise from dipole-allowed O $1s$ transitions targeting unoccupied states with O $2p$--TM $3d$ hybridization, and are therefore sensitive to changes in metal--oxygen bonding and local coordination accompanying cathode degradation~\cite{degr+89prb}. However, the interpretation of experimental spectra is often hindered by intrinsic excitation lifetimes and sample inhomogeneities, demanding insight from theory~\cite{vorw+17jpcl,wibo+25jpcc,garc+25jpcl}. First-principles calculations based on density functional theory (DFT) and many-body perturbation theory (MBPT) have emerged as essential complements to experimental observations for resolving electronic structure and producing the spectral fingerprints of complex materials~\cite{cocc+16prb,vorw+17jpcl,cocc20pssrrl,mach+23ic,duar+25rsca,xu+25pccp}. Solving the Bethe-Salpeter equation (BSE) for core-level excitations in an all-electron framework~\cite{lask-blah10prb,vorw+17prb,vorw+19es} provides an accurate, parameter-free description of the many-body effects ruling X-ray excitations.

Despite their successes, first-principles methods face a fundamental hurdle when simulating real-world samples: the input structural model. While assuming pristine, single-phase crystal structures is the standard procedure for characterizing conventional semiconductors~\cite{lask-blah10prb}, the simulation of spectra of battery materials is inherently more complex due to coexisting phases and structural disorder in the sample. Libraries collecting the spectral fingerprints of reference materials have been successfully used to decode measurements, allowing for the identification of hidden or overlapping structural components and local coordination~\cite{seo+16natchem} and the identification of otherwise inaccessible metastable intermediates~\cite{dong+26cej}. Moreover, these theoretical databases provide a crucial foundation for data-driven and machine learning models~\cite{khar+25prm} for the automated screening of phase transitions and secondary phases at an unprecedented level of accuracy~\cite{sha+22npjcm}.

In this paper, we integrate MBPT-calculated XAS spectra with an efficient, DFT-based high-throughput screening to decipher the O K-edge spectra of NCM-811 degradation products using local structural proxies. We first identify the core-level fingerprints of eight nominal reference oxides with layered, spinel, and rock-salt lattices, and subsequently extend our analysis to 38 theoretically predicted binary polymorphs with general formula \ce{MO} and \ce{M3O4} (M = Ni, Co, Mn), resulting from database mining and high-throughput filtering. By establishing the O $p$-contributions of the projected density of states (PDOS) as a reliable descriptor across the explored structural pool, we identify how local structural variations contribute to the macroscopic experimental signatures of nominal references. This combined computational approach provides a robust framework for identifying degradation products in batteries with Ni-rich cathodes such as NCM-811.


\section{Methods}
\subsection{Theoretical Background and Computational Details}
The computational results presented in this work are obtained from first principles in the framework of DFT and MBPT.
To compute core-level excitations, we solve the BSE in the all-electron full-potential framework provided by the \texttt{exciting} code~\cite{gula+14jpcm,vorw+17prb,vorw+19es}. The BSE problem is mapped into the two-particle eigenvalue equation,
\begin{equation}\label{eq:bse}
  \sum_{c'u'\textbf{k'}}\hat{H}^{BSE}_{cu\textbf{k},c'u'\textbf{k'}}A^{\lambda}_{c'u'\textbf{k'}}=E^{\lambda}A^{\lambda}_{cu\textbf{k}},
\end{equation}
where the indices $c$ and $u$ indicate core and fully unoccupied states, respectively. The BSE Hamiltonian is the sum of three terms: $\hat{H}^{BSE} = \hat{H}^{diag} + \hat{H}^{dir} + \hat{H}^x$. The diagonal term ($\hat{H}^{diag}$) accounts for single-particle vertical transitions, while the direct ($\hat{H}^{dir}$) and exchange ($\hat{H}^x$) terms incorporate the direct, statically screened electron-hole Coulomb attraction and the repulsive exchange interaction between them, respectively. Neglecting the two Coulomb parts of the BSE Hamiltonian corresponds to the so-called independent-particle approximation (IPA), in which excitonic effects are neglected.

The eigenvalues of Eq.~\eqref{eq:bse}, $E^{\lambda}$, correspond to the excitation energies, while the eigenvectors $A^{\lambda}$ carry information about the character of the excitations. They both enter the expression of the imaginary part of the macroscopic dielectric function,
\begin{equation}\label{eq:epsilon}
  \Im{\varepsilon_{M}}=\frac{8\pi^{2}}{\Omega}\sum_{\lambda}\mid\textbf{t}_{\lambda}\mid^{2}\delta(\omega-E_{\lambda}),
\end{equation}
which is used to represent the absorption spectrum of the material. Beyond the unit-cell volume $\Omega$, Eq.~\eqref{eq:epsilon} includes the square moduli of the transition coefficients 
\begin{equation}\label{eq:t}
  \mathbf{t}_{\lambda}=\sum_{cu\mathbf{k}}A^{\lambda}_{cu\mathbf{k}}\frac{\bra{c}\hat{\mathbf{p}}\ket{u\mathbf{\mathbf{k}}}}{\epsilon_{u\mathbf{k}} - \epsilon_{c}+\Delta},
\end{equation}
where the energy difference in the denominator is supplemented by a scissors operator $\Delta$ determined from experiments to ensure an accurate onset for the calculated spectra.

All-electron DFT and MBPT calculations are performed with the \textit{ab initio} code \texttt{exciting}~\cite{gula+14jpcm}. The electronic structure of the materials is computed using the Perdew–Burke–Ernzerhof (PBE) exchange-correlation functional~\cite{pbe}. Input structures taken from Materials Project~\cite{jain+13aplm} were relaxed with an interatomic forces threshold of 1~meV/\AA{}. Brillouin zone sampling using Monkhorst-Pack grids was tailored to each system and detailed in Table~S1. Likewise, the muffin-tin radii were optimized for each species and compound (Table~S1), adopting a plane-wave cutoff of $R_{MT}G_{MAX}=8$ (reduced to 7 for \ce{LiMnO2}) for basis-set convergence (see Ref.~\citenum{gula+14jpcm} for detailed explanations of these parameters).
To solve the BSE, $\Gamma$-shifted $\mathbf{k}$-grids were used (Table~S1) along with unoccupied states spanning 30~eV above the Fermi level. The scissors operator $\Delta$ (Eq.~\ref{eq:t}), ranging between 21.4 and 26.6~eV in the considered material pool (Table~S1) was adjusted to match the experimental onset of each system. We emphasize that adopting a system-dependent scissor-shift when comparing BSE-calculated spectra is standard practice in the community~\cite{vorw+17prb,vorw+19es} when comparing the results against their respective experimental reference datasets.

To explore the influence of polymorphism and magnetic ordering on the spectral fingerprints of the considered oxides, we complemented the MBPT analysis with a high-throughput DFT screening based on the plane-wave pseudopotential code \texttt{Quantum ESPRESSO}~\cite{Giannozzi2017} and automated using the \texttt{AiiDA} infrastructure~\cite{Uhrin2021,Huber2020}.
Magnetic configurations were enumerated using the \texttt{enumlib} library~\cite{Hart2012} implemented in \texttt{aim2dat}~\cite{sassnick2026aim2dat}. Following structural relaxation and magnetic ground-state identification, we computed the PDOS for 38 distinct polymorphs and used it as a primary descriptor to interpret experimental XAS features, shedding light on the discrepancies emerging in selected BSE results for single phases. Details of the automated workflow and analysis are provided in Figure~S2.

The high-throughput screening was performed using the PBE functional~\cite{pbe} for consistency with the all-electron DFT calculations underlying the BSE analysis. Despite the well-known limitations of semilocal functionals in underestimating band gaps of semiconducting oxides and spuriously shifting the energies of $3d$ states relative to oxygen contributions~\cite{dixi+17jpcc}, the relative bandwidth and morphology of the oxygen $2p$ PDOS features remain robust across different exchange-correlation functionals~\cite{zira+22ps} (further details in Figures~S6-S8 of the Supporting Information). As such, the PBE-calculated PDOS provides a reliable fingerprint for structure-spectral relations. Additional comparison between BSE results and independent-particle-approximation counterparts confirms that excitonic effects captured only by the former approach induce only a rigid redshift of a few hundred meV in the excitation energies without altering the qualitative peak distribution.

\subsection{Experimental Methods}
O K-edge partial fluorescence yield (PFY) and total electron yield (TEY) soft XAS spectra were obtained at the Advanced Light Source, beamline 8.0.1, at the iRIXS endstation~\cite{qiao+17rsi}. While the \ce{NiO} (99.995\% trace metal basis; Sigma Aldrich, USA),  \ce{MnO} ($\geq$ 99.99\% trace metal basis; Alfa Aesar, USA), \ce{CoO} ($\geq$ 99.995\% trace metal basis; Alfa Aesar, USA), \ce{Co3O4} ($\geq$ 99.9985\% trace metal basis; Alfa Aesar, USA), \ce{Mn3O4} (99.9\% trace metal basis; Sigma Aldrich, USA) powder samples were mounted in an air atmosphere and introduced into the ultrahigh vacuum (UHV) iRIXS analysis chamber (base pressure $<10^{-9}$~mbar) via a load-lock, the LiCoO$_2$ (99.5\% trace metal basis; Sigma Aldrich, USA), LiMnO$_2$ ($\geq98$\% trace metal basis; Sigma Aldrich, USA), LiNiO$_2$ ($\geq98$\% trace metal basis; Sigma Aldrich, USA) powder samples were mounted in an Ar-filled glovebox and transferred into the ultrahigh vacuum (UHV) analysis main chamber (base pressure $<$10$^{-9}$~mbar) using a transfer suitcase to avoid air exposure. The entrance and exit slit of the monochromator were set to 50~$\mu$m, resulting in a beamline resolution of 0.4~eV and a beam size on the sample of 25~$\mu$m in diameter. 

To minimize X-ray irradiation-induced damage, the samples were continuously moved up and down $\pm1$~mm relative to the focused beam during data collection. The photon energy was calibrated using a TiO$_2$ reference sample~\cite{lusv+98ss}. For the partial fluorescence yield X-ray absorption spectroscopy (PFY-XAS) data, O K-edge Resonant Inelastic X-ray Scattering (RIXS) maps were collected by means of measuring X-ray emission spectra (XES) in an excitation energy range between 520 and 525.5~eV using an energy step size of 0.5~eV, in an excitation energy range between 526 and 533~eV using an energy step size of 0.2~eV and in an excitation energy range between 533.5 and 550~eV using an energy step size of 0.5~eV and a measurement time of 20~s by the hrRIXS spectrometer~\cite{qiao+17rsi}. The RIXS maps were visualized using an \texttt{Igor} code written for BL 8.0.1~\cite{lusv+98ss}. The emission energy was calibrated based on the elastic peak line, coinciding with the photon energy, visible in the RIXS map. The O K-edge PFY-XAS spectra were then obtained by integrating the RIXS map in the direction of the emission energy between 518 and 536~eV and plotting this PFY intensity over the excitation energy. The TEY-XAS data were measured simultaneously with PFY by collecting the drain current of the sample using a Keithley 6517B picoampermeter.

\section{Results and Discussion}
\begin{figure}[h]
  \centering
  \includegraphics[width=\textwidth]{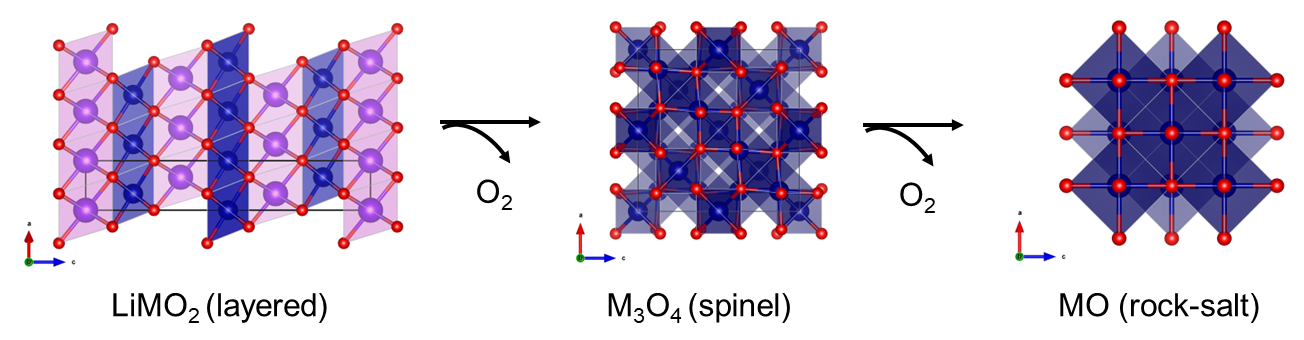}
  \caption{Ball-and-stick representation of the crystal structures of layered lithiated transition-metal (M = Co, Mn, Ni) oxides (left), delithiated spinel (middle), and rock-salt structures (right). These idealized lattices represent the primary structural models for the nominal reference oxides formed following \ce{O2} release. Transition metal atoms are depicted in blue, O atoms in red, and Li atoms in violet.}
  \label{fgr:systems}
\end{figure}

To investigate and rationalize the spectral fingerprints of aging mechanisms in NCM-811 cathodes~\cite{debi+17jpcc, jia+22nl}, we consider a suite of transition-metal oxides including layered ($R\bar{3}m$), spinel ($Fd\bar{3}m$), and rock-salt ($Fm\bar{3}m$) phases~\cite{kond+17jpcc, zou+18acsel, hua+19natcom, chen+21nano, gan+23nr}, representing characteristic degradation products of \ce{LiNi_{0.8}Co_{0.1}Mn_{0.1}O2}~\cite{hu+21acsami}. As illustrated in Figure~\ref{fgr:systems}, these systems capture the structural evolution from the pristine lithiated layered oxides (\ce{LiMO2}) to the lower-symmetry spinel (\ce{M3O4}) and rock-salt (\ce{MO}) phases that form upon oxygen loss and metal-ion migration. 

\subsection{Spectroscopic Characterization of Reference Oxides}

\begin{figure}[h]
  \centering
  \includegraphics[width=\textwidth]{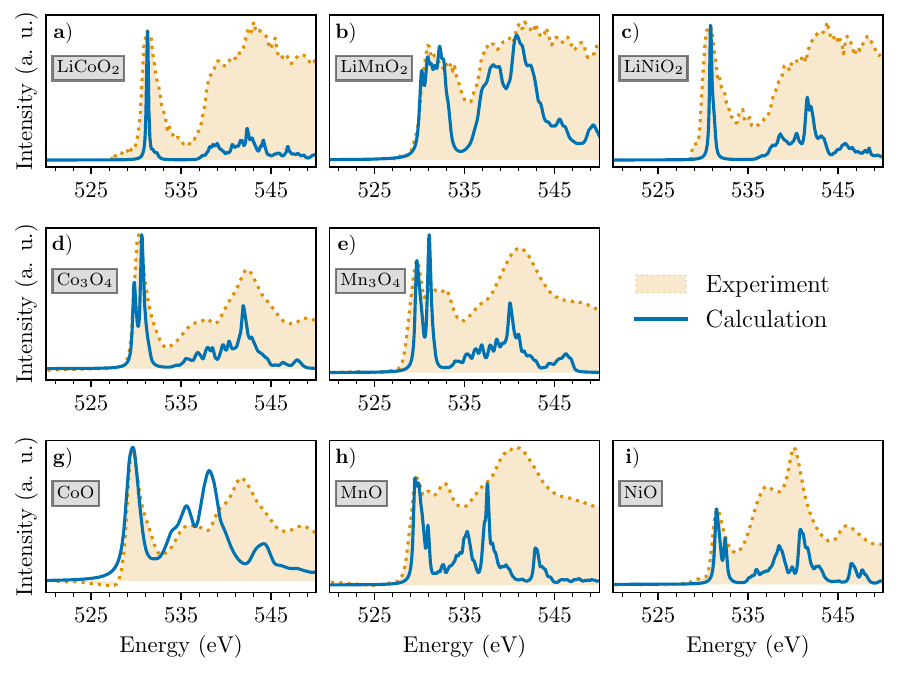}
  \caption{X-ray absorption spectra of a) \ce{LiCoO2}, b) \ce{LiMnO2}, c) \ce{LiNiO2}, d) \ce{Co3O4}, e) \ce{Mn3O4}, f) \ce{CoO}, g) \ce{MnO}, and h) \ce{NiO}. The reference experimental data represent PFY-XAS (a-c) and TEY-XAS (d-h) data. A Lorentzian broadening of 110~meV and 170~meV is used for the computed spectra in panels (a-c) and (d-h), respectively, to allow for a fair comparison to the experimental XAS data.
  }
  \label{fgr:BSEvsEXP}
\end{figure}

We begin our analysis by evaluating the O K-edge X-ray absorption spectra for the eight reference systems, comparing BSE-computed with experimentally derived data (Figure~\ref{fgr:BSEvsEXP}).
All spectra are dominated by an intense resonance at the absorption onset originating from O $1s \rightarrow 2p$ transitions into the bottom of the conduction band. While the experiments were able to resolve a single peak in all considered samples, the calculations reveal a more complex spectral structure, particularly for \ce{Co3O4} and \ce{Mn3O4}, where two energetically close resonances dominate the absorption (Figure~\ref{fgr:BSEvsEXP}d,e). Similar features also appear in the computed spectra of \ce{MnO} and \ce{NiO} (Figure~\ref{fgr:BSEvsEXP}g,h). However, in these cases, corresponding shoulders are visible in the experiment, too. 

At higher energies, all spectra are dominated by transitions to higher-lying unoccupied states, giving rise to broad absorption in the measurements, generally reproduced by all simulations. However, while the experiment-theory agreement varies across reference compounds, the spectra of the binary rocksalt oxides are more challenging to reproduce from first principles. In these highly symmetric isotropic lattices, the peak widths and relative intensity ratios are highly susceptible to local structural distortions and the specific representation of the local electronic environment. Consequently, the discrepancies observed in the rock-salt spectra suggest that the experimental samples likely encompass local symmetry-breaking variations that are absent in a perfectly symmetric, single-phase bulk model.

Absolute electronic energies can be sensitive to the choice of the exchange-correlation functional. While a Hubbard $U$ correction is routinely used to mitigate self-interaction errors and improve localized electronic descriptions of transition-metal oxides~\cite{okum+12jmc,Chakraborty2018,coro+26cm}, incorporating on-site parameters introduces system- and coordination-dependent variations across different structural motifs~\cite{moor+24prm}, potentially obscuring the underlying physical trends across a broad library of candidate polymorphs. To maintain consistency throughout this analysis and with previous work~\cite{reen+25sr,duar-cocc22jpcc}, we use the semi-local PBE functional across all materials. In doing so, we acknowledge the quantitative energy shifts inherent to semi-local approximations and instead focus on the topological robustness of the O $2p$ PDOS morphology (further details in Section S6 of the Supporting Information; see Figures~S6-S8). A thorough comparison with the literature~\cite{gill-robe13jpcm,seo+15prb} reveals that the relative peak distribution and hybridization patterns, which are relevant to the spectroscopic fingerprinting, are qualitatively preserved across functionals. This consistency ensures that our comparative structural analysis remains robust against the absolute energetic discrepancies typically encountered in single-phase bulk models.

To dissect these spectral features systematically, we first focus on the layered oxide series. Among them, the calculated spectra of trigonal \ce{LiCoO2} and \ce{LiNiO2} exhibit striking similarities (Figure~\ref{fgr:BSEvsEXP}a,c). Both are characterized by a sharp onset around 530~eV, followed by a broader absorption feature at higher energies, starting from approximately 536~eV. Despite these analogies, the calculated relative intensities differ: in \ce{LiCoO2}, the first peak is roughly four times more intense than the higher-energy maxima, whereas in \ce{LiNiO2}, a distinct local maximum appears at $\sim$542~eV with approximately half the intensity of the main peak. In contrast, the experimental spectrum of \ce{LiMnO2} (Figure~\ref{fgr:BSEvsEXP}b) is dominated by a broader onset spanning almost 5~eV (from 529~eV to 534~eV), followed by a region of equally intense absorption with discernible peaks at 537~eV and 541~eV. Both spectral regions are well reproduced by our BSE calculations. We note that the experimental intensities above 535~eV are further enhanced by the transitions into continuum states, which manifests as a rising background step. As our BSE framework focuses on the excitonic transitions in the pre-edge and absorption onset, this continuum contribution is not explicitly modeled, magnifying the intensity differences at higher energies.

\begin{figure}[htb!]
  \centering
  \includegraphics[width=\textwidth]{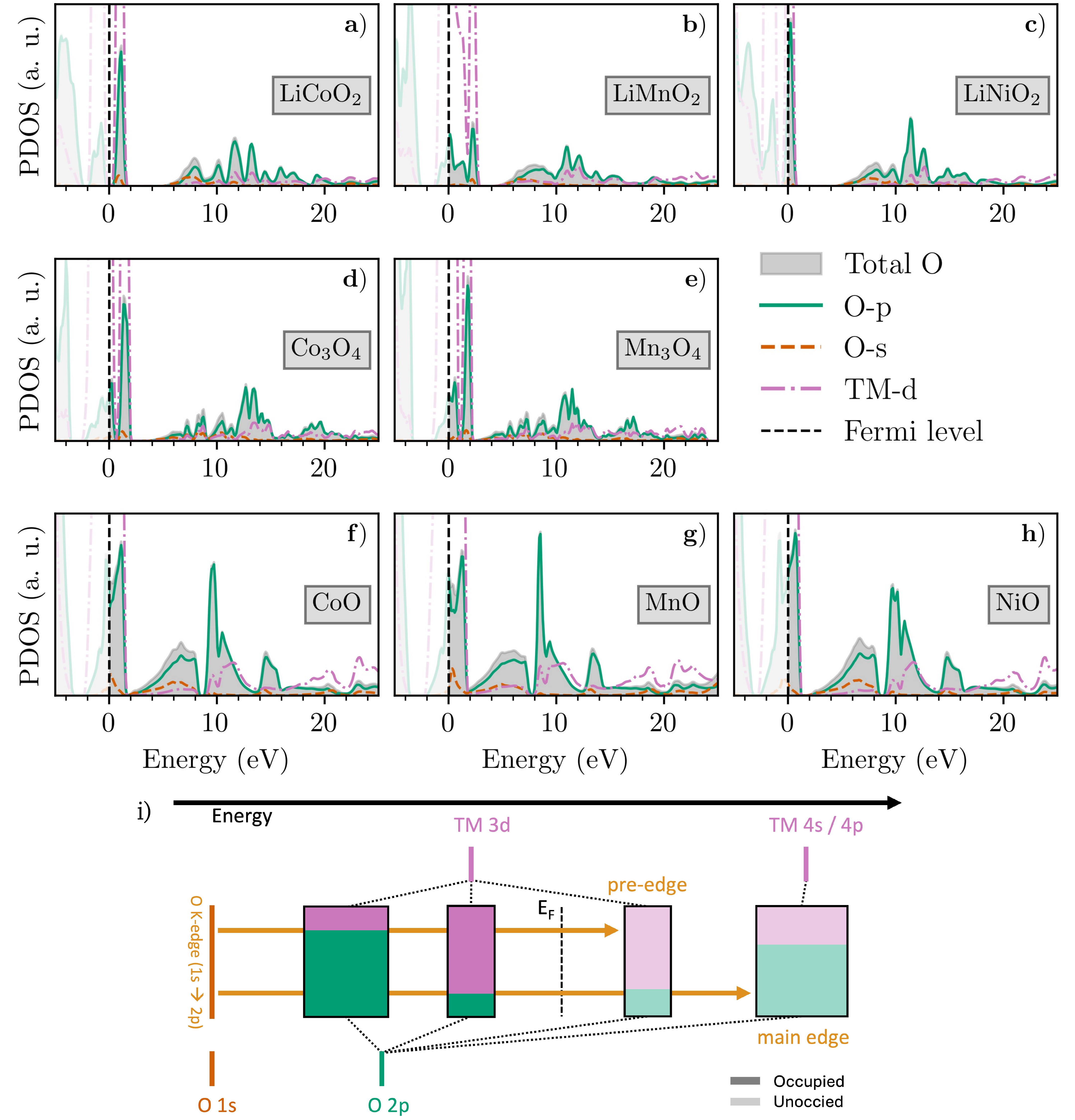}
  \caption{PDOS of a) \ce{LiCoO2}, b) \ce{LiMnO2}, c) \ce{LiNiO2}, d) \ce{Co3O4}, e) \ce{Mn3O4}, f) \ce{CoO}, g) \ce{MnO}, and h) \ce{NiO}. Oxygen $s$- and $p$-states are highlighted (green solid and orange dashed lines, respectively) compared to the total O contribution (gray shaded area) and transition-metal (TM) $d$-states (pink dotted-dashed line). The Fermi level is marked by a dashed black bar. i) Schematic representation of the O $1s$ transitions targeting unoccupied hybridized $2p--3d$ states relative to the Fermi level ($E_F$).
  }
  \label{fgr:Opdos}
\end{figure}

Overall, the BSE results successfully reproduce the primary O K-edge PFY-XAS measurements of the layered oxides (Figure~\ref{fgr:BSEvsEXP}a-c), maintaining a consistent distribution of spectral weight. The observed enhancement in the experimental yield at high energies is ascribed to the contributions of continuum states and the larger intrinsic broadening typical of high-energy experimental features. The underestimated width of the first peak around 530~eV appears as a systematic, minor discrepancy across all calculated spectra. We emphasize that in our computational framework, the peak broadening is controlled by a Lorentzian function added in post-processing. Crucially, the pre-edge shoulder at 529~eV in \ce{LiNiO2} (Figure~\ref{fgr:BSEvsEXP}c, red arrow) is a well-documented signature of partial lithium deintercalation (\ce{Li_{1-x}NiO2})~\cite{ushi+01jps}. This feature stems from spontaneous surface delithiation driven by the high formation enthalpies of parasitic surface phases (e.g., \ce{Li2O2} and \ce{Li2CO3}), even under inert glovebox conditions, which introduces O $2p$ hole states below the formal conduction edge. In \ce{LiMnO2} (Figure~\ref{fgr:BSEvsEXP}b), the calculated spectrum features a subtle intensity redistribution: while the experimental yield peaks at the first resonance ($\sim$531 eV), the BSE calculations predict a slightly higher weight for the subsequent feature at $\sim$532~eV. This variation points to the high sensitivity of orbital hybridization to the local symmetry breaking and coordination environments. To gain a deeper insight into the origin of these spectral features, it is instructive to examine the electronic structure via the O $p$-contributions to the PDOS for all considered binary oxides (Figure~\ref{fgr:Opdos}a-h) supplemented by a schematic illustration of oxygen $1s$ transitions to unoccupied hybridized $2p-3d$ states relative to the Fermi level (Figure~\ref{fgr:Opdos}a-h).

In this analysis, we exploit the topological robustness of the semi-local electronic structure to interpret these spectral trends. While the PBE functional systematically underestimates the absolute band gaps of transition-metal oxides~\cite{zira+22ps}, a comparison between the unoccupied PDOS profiles obtained via PBE and PBE+$U$ protocols (Figure~S5) confirms that the relative distribution and morphology of the unoccupied O $p$-states remain qualitatively unaltered. As shown in Figure~\ref{fgr:Opdos}, the electronic structure of \ce{LiMnO2} differs strikingly from its Co and Ni counterparts. In \ce{LiCoO2} and \ce{LiNiO2} (Figure~\ref{fgr:Opdos}a,c), the O $p$-states are distinctly partitioned: broader bands characterize the valence region, while a sharp, isolated maximum dominates the bottom of the conduction band. In contrast, the unoccupied O $p$-state manifold in \ce{LiMnO2} (Figure~\ref{fgr:Opdos}b) is energetically downshifted toward the valence edge (vertical dashed line in Figure~\ref{fgr:Opdos}), resulting in a significantly higher density of unoccupied states immediately above the absorption threshold. This high density of available states at the conduction edge provides a clear physical explanation for the broad, multi-peaked onset observed in its experimental XAS profile (Figure~\ref{fgr:BSEvsEXP}b).

This energetic redistribution of the electronic states directly explains the enhanced oscillator strength of the higher-energy maxima in \ce{LiMnO2} relative to \ce{LiCoO2} and \ce{LiNiO2}. Importantly, the BSE calculations successfully capture these relative intensities across the series of layered oxides (Figure~\ref{fgr:BSEvsEXP}a-c), validating the necessity of accounting for explicit electron-hole correlations in the \textit{ab initio} simulations. We note that the characteristic step-like intensity increase at higher energies in the experimental spectra can be modeled via an energy-dependent Lorentzian broadening function to account for the decreasing quasiparticle lifetime of high-energy excitations~\cite{vorw+17prb,vorw+17jpcl}. However, employing a constant broadening is sufficient in this analysis, which focuses strictly on primary peak assignments and structural fingerprinting.

Next, we turn to the spinel structures with the general formula \ce{M3O4}. Here, we consider exclusively \ce{Co3O4} and \ce{Mn3O4}, as bulk \ce{Ni3O4} is thermodynamically unstable under ambient conditions~\cite{kubo-hu11iecr}. A direct comparison between their XAS profiles (Figure~\ref{fgr:BSEvsEXP}d,e) reveals a significantly broader absorption onset in \ce{Mn3O4} than in \ce{Co3O4}. The BSE calculations on  \ce{Co3O4} resolve two closely spaced excitations within the lowest-energy peak, centered at approximately 529.5~eV and 531~eV. The lower-energy resonance is weaker than the primary peak, which excellently reproduces the experimental intensity distribution. Conversely, for \ce{Mn3O4} (Figure~\ref{fgr:BSEvsEXP}e), the calculated spectral width at the onset is slightly underestimated compared to the experiment, yielding a double-peak structure with maxima at 530~eV and 531~eV. Above 532~eV, the simulated intensity drops, whereas the experimental data exhibit a 2~eV-wide plateau, followed by a relative minimum at 535~eV and a subsequent rise toward a major maximum at 541~eV, which is well mirrored by the BSE results. An analogous high-energy feature is also discernible in the spectrum of \ce{Co3O4}, shifted slightly higher to around 542--543~eV.

To trace the physical origin of these features, we inspect the corresponding PDOS profiles. In the case of \ce{Co3O4} (Figure~\ref{fgr:Opdos}d), a sharp manifold of unoccupied O $2p$ states hybridized with transition-metal levels is positioned immediately above the valence-band maximum (VBM). These empty states are the target for the O $1s \rightarrow 2p$ transitions that form the low-energy onset shoulder in the XAS data (Figure~\ref{fgr:BSEvsEXP}d). A similar electronic mechanism occurs in \ce{Mn3O4} (Figure~\ref{fgr:Opdos}e), where a notably broader energetic distribution of unoccupied O $2p$ states spans the conduction band edge, leading to the increased width of the first experimental absorption peak. The slight underestimation of this onset width in the BSE simulation is likely a consequence of the complex structural landscape of \ce{Mn3O4}, which experimentally accommodates a cooperative Jahn-Teller tetragonal distortion that can break local symmetries beyond the idealized spinel model~\cite{tack+07prb,keme+14prb}. For both compounds, the primary absorption features within the first few electronvolts of the threshold originate from transitions into these heavily hybridized, low-lying unoccupied O $p$-sets at the bottom of the conduction band landscape.

The spectra of the rock-salt oxides (general formula \ce{MO}) are characterized by a distinct peak at the absorption onset, followed by a broader, structured absorption region (Figure~\ref{fgr:BSEvsEXP}f-h). The best agreement between theory and experiment is achieved for \ce{NiO} (Figure~\ref{fgr:BSEvsEXP}h), for which the relative peak energies and oscillator strengths match almost perfectly when accounting for the linearly increasing experimental broadening that is omitted in the calculations. Remarkably, the higher-energy shoulder of the primary peak is correctly captured as well. Very good agreement between the calculations and experiments is also obtained for \ce{CoO} (Figure~\ref{fgr:BSEvsEXP}f), particularly at the absorption threshold, where the first maximum at $\sim$530~eV is almost identically reproduced by the BSE results. At higher energies, the broad peak structure between 533~eV and 536~eV is well reproduced by the calculation, while the subsequent peak, appearing in the measurement at 541.5~eV, is predicted at 538~eV from first principles. A similar shift occurs for the high-energy shoulder, which is visible in the range 543--545~eV in the calculated spectrum and between 548--550~eV in the measurement (Figure~\ref{fgr:BSEvsEXP}f). The origin of this systematic shift can be attributed to well-known limitations of the semi-local PBE functional in describing the highly localized self-interaction corrections and spatial splitting of hybridized transition-metal-oxygen bands~\cite{zira+22ps,dixi+17jpcc} (Figure~S5). Finally, for \ce{MnO}, the discrepancy between the BSE results and the measurement is most pronounced (Figure~\ref{fgr:BSEvsEXP}g). While the first peak at 530~eV is correctly captured, including the weaker high-energy shoulder at approximately 531~eV, the higher-energy resonances, reproduced as distinct peaks in the calculation, appear as a broad, damped, and rather flat signal in the experiment (532--540~eV). Similar to \ce{CoO}, the maximum recorded around 540~eV is systematically underestimated by about 5~eV in the BSE spectrum, where it aligns with a sharper calculated resonance at 538.5~eV. This deviation from the sharp, discrete peaks predicted for the idealized lattice suggests the emergence of distortion or local structural disorder in the experimental sample.

From the analysis of the PDOS for these compounds (Figure~\ref{fgr:Opdos}f-h), we can assign the first peak in the XAS data for the rock-salt oxides to transitions into unoccupied O $2p$ states hybridized at the conduction band edge immediately above the VBM. The higher-energy maxima correspond to excitations targeting higher levels in the unoccupied manifold. The imperfect quantitative description of these high-energy features in the BSE spectra can be ascribed to the underlying description of transition-metal $d$-states within the semi-local PBE approximation. Interestingly, these electronic artifacts are less prominent in the electronic structure of \ce{NiO}, as demonstrated by the excellent agreement between the BSE and experimental results (Figure~\ref{fgr:BSEvsEXP}h). To gain deeper insights into the outlined inconsistencies between calculations and measurements, we transition in the following from the high-accuracy framework of MBPT to a high-throughput screening of the structural landscape.

\subsection{High-Throughput Analysis}
The BSE analysis of the binary oxides, which serve as baseline models for potential degradation products of NCM-811, revealed systematic discrepancies compared to experiment, specifically regarding peak shapes and oscillator strength distributions (Figures~\ref{fgr:BSEvsEXP}f–h). While the known limitations of PBE in describing metal-oxygen hybridization~\cite{li+13jctc,molt-kepp19chpch} play a role, the assumption of an idealized single-crystal bulk lattice in \textit{ab initio} simulations exerts an even more significant influence. This structural idealization fails to capture the intrinsic local variations present in nominal reference compounds and becomes even less applicable to actual electrochemical degradation layers, which naturally encompass reconstructed, strained, and disordered domains. To bridge this gap, we adopt a high-throughput diagnostic framework that moves beyond single-phase bulk models, using polymorphic structural proxies to understand how local coordination heterogeneity shapes experimental XAS profiles. 

To assess the impact of structural variations on calculated core-level excitations, we employ a high-throughput DFT approach. While the rock-salt phase represents the thermodynamic ground state under ambient conditions~\cite{shur20jes,li+24mre}, cycled battery cathodes systematically exhibit highly reconstructed, disordered surface layers where nominal lattice symmetry is broken. To capture proxies for local lattice strain and coordination environments, we screened a pool of 104 compounds with the \ce{MO} composition ($\text{M = Co, Ni, Mn}$), queried from the Materials Project~\cite{jain+13aplm} and the Open Quantum Materials Database~\cite{Kirklin2015} (further details in the SI). 
After performing a uniqueness analysis based on structural similarity using the F-fingerprint method~\cite{Oganov2009} and excluding non-converged calculations, we obtained a set of 68 candidate structures. Following full structural relaxation and subsequent uniqueness verification, this set was refined to 38 distinct stable compounds (Figure~S3). This screened set spans a broad structural parameter space, effectively covering variable TM-O coordination numbers (ranging from 4- to 6-fold) and anisotropic bond-length distortions (Figure~S4). To complement this structural filtering, we clustered the resulting electronic density profiles to ensure uniquely representative electronic environments across the different oxides~\cite{reen+25sr}.

We emphasize that these candidate polymorphs are not assumed to coexist as long-range, thermodynamically stable bulk phases within a single sample, nor are they intended to serve as exhaustive structural replicas of complex defective networks, such as explicit point vacancies or random amorphous boundaries. Rather, they serve as structural proxies, namely local geometric and electronic descriptors that capture specific motifs of coordination distortion, anisotropic strain, and symmetry breaking. By establishing the unoccupied O $p$-PDOS as a computationally accessible descriptor for the energy distribution of available target states, we scan our library of oxide polymorphs to diagnose how specific local symmetry-broken configurations account for spectral broadening, feature shifts, and shoulder formation observed in experiment.

\begin{figure}[t!]
  \centering
  \includegraphics[width=\textwidth]{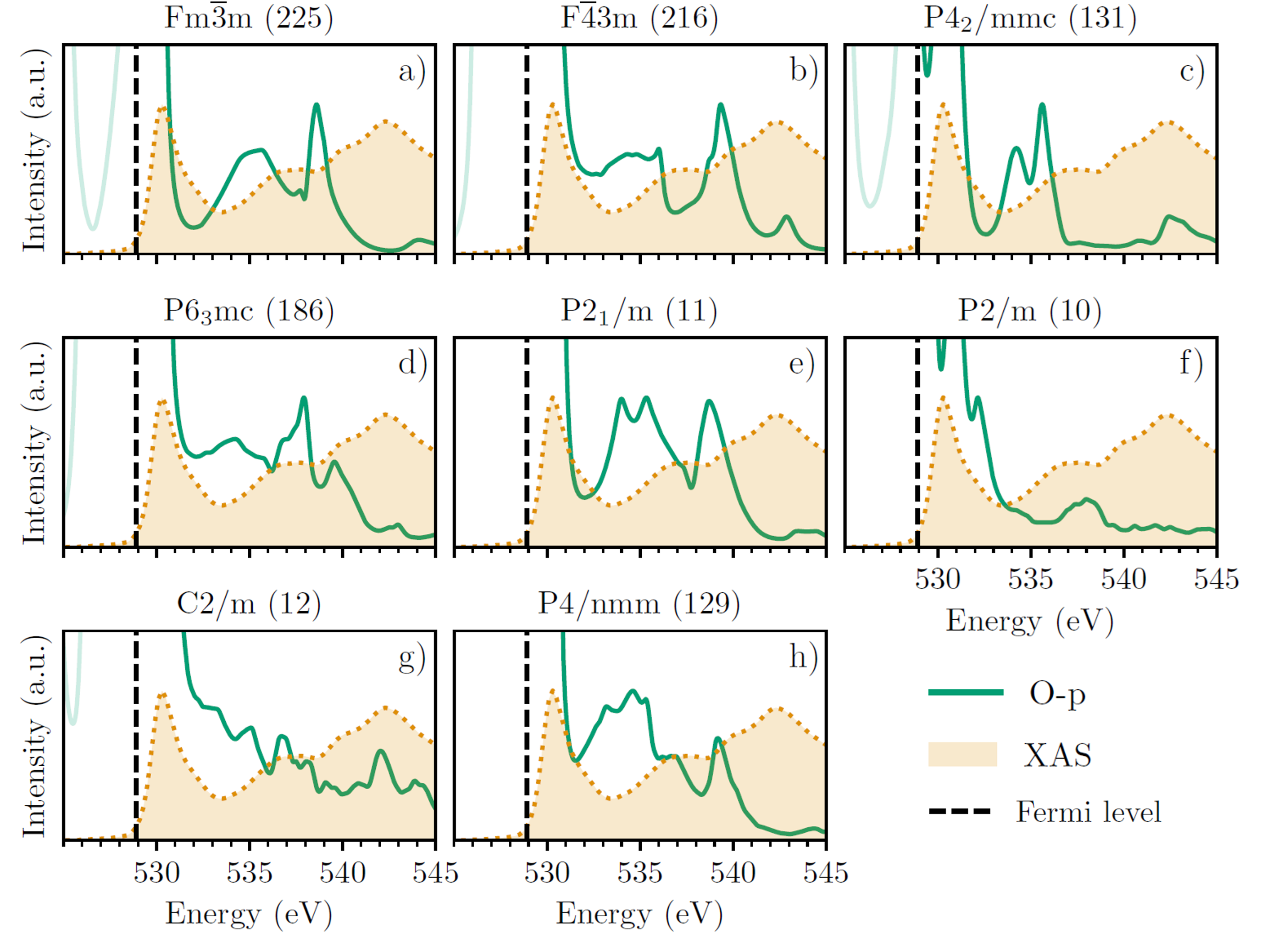}
\caption{Comparison between the calculated unoccupied O $2p$ PDOS for CoO polymorphs (green) and the experimental TEY-XAS spectrum of nominal CoO (red): \textbf{a)}~$Fm\bar{3}m$ (225), \textbf{b)}~$F\bar{4}3m$ (216), \textbf{c)}~$P4_2/mmc$ (131), \textbf{d)}~$P6_3mc$ (186), \textbf{e)}~$P2_1/m$ (11), \textbf{f)}~$P2/m$ (10), \textbf{g)}~$C2/m$ (12), and \textbf{h)}~$P4/nmm$ (129). A Lorentzian broadening of 170~meV is applied to the PDOS curves. All spectra are aligned relative to the calculated Fermi energy by fixing the first conduction peak of the $Fm\bar{3}m$ phase to the experimental onset.}
  \label{fgr:ht_CoO_all}
\end{figure}

We begin our analysis with the results for \ce{CoO}. While the ground-state rock-salt phase (Figure~\ref{fgr:ht_CoO_all}a) demonstrates good agreement with the experimental features, other variants exhibit PDOS characteristics highly compatible with the nominal \ce{CoO} reference spectrum (Figure~\ref{fgr:ht_CoO_all}d,f). Notably, the tetrahedrally coordinated zinc-blende phase (Figure~\ref{fgr:ht_CoO_all}b) shows excellent agreement in terms of the primary peak positions, yielding a spectral match comparable to the pristine rock-salt phase. Finally, while the alignment is less precise for the 535~eV feature in the tetragonal polymorph (Figure~\ref{fgr:ht_CoO_all}c), its first peak and high-energy feature at 543~eV track the experimental data well. These distinct variations, i.e., the differing alignment quality across specific spectral regions, highlight that from a spectroscopic perspective, an array of structurally diverse local environments could form during battery operation, collectively contributing to the observed macro-scale experimental profiles.

\begin{figure}[t]
  \centering
  \includegraphics[width=\textwidth]{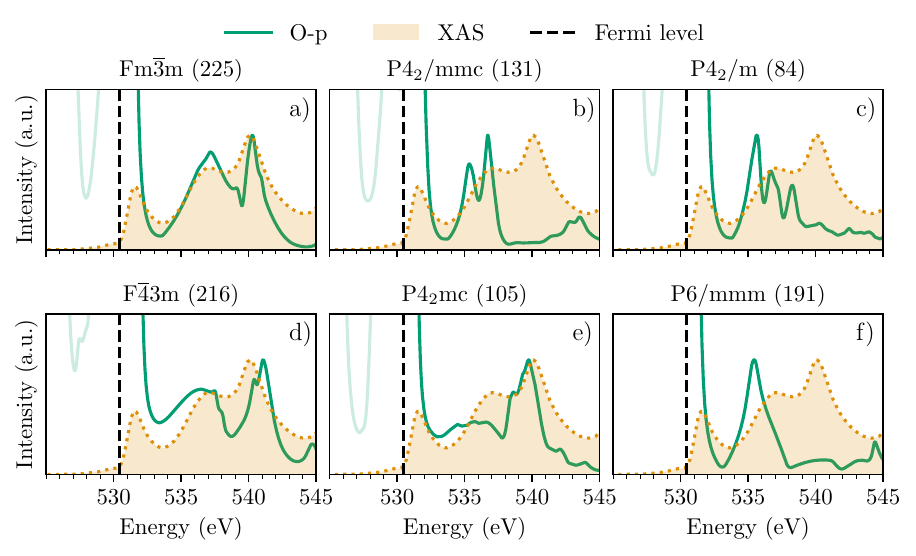}
\caption{Comparison between the calculated unoccupied O $2p$ PDOS for NiO polymorphs (green) and the experimental TEY-XAS spectrum of nominal NiO (red): \textbf{a)}~$Fm\bar{3}m$ (225), \textbf{b)}~$P4_2/mmc$ (131), \textbf{c)}~$P4_2/m$ (84), \textbf{d)}~$P6_3/mmc$ (194), \textbf{e)}~$F\bar{4}3m$ (216), and \textbf{f)}~$P4_2mc$ (105). A Lorentzian broadening of 170~meV is applied to the PDOS curves. All spectra are aligned relative to the calculated Fermi energy by fixing the first conduction peak of the $Fm\bar{3}m$ phase to the experimental onset.}
  \label{fgr:ht_NiO_all}
\end{figure}

Next, we examine the structural library of \ce{NiO} polymorphs. As shown in Figure~\ref{fgr:ht_NiO_all}a, the unoccupied O $p$-PDOS of the ground-state rock-salt structure is in excellent agreement with the experimental XAS data of the nominal \ce{NiO} reference. Interestingly, alternative \ce{NiO} phases, including the tetrahedrally coordinated zinc-blende structure (Figure~\ref{fgr:ht_NiO_all}e) and distinct higher-energy cubic variants (Figure~\ref{fgr:ht_NiO_all}d), reproduce the core spectroscopic fingerprints of the O $2p$ target states, such as the prominent resonance feature at 545~eV. While these non-equilibrium phases are typically stabilized only under extreme mechanical/thermodynamic constraints or as nanostructured thin-film surface reconstructions~\cite{eto+00prb,gavr+23comphys}, their close spectral alignment with the experimental XAS profile suggests potential contributions from local coordination motifs that deviate from the idealized bulk ground state. These findings demonstrate that the primary spectroscopic signatures of \ce{NiO}-based degradation products remain remarkably robust against variations in long-range crystal symmetry within the same nominal composition. Conversely, the more complex, heavily distorted local environments characteristic of \ce{MnO} necessitate a more nuanced structural interpretation, as detailed below.

\begin{figure}[h!]
  \centering
  \includegraphics[width=\textwidth]{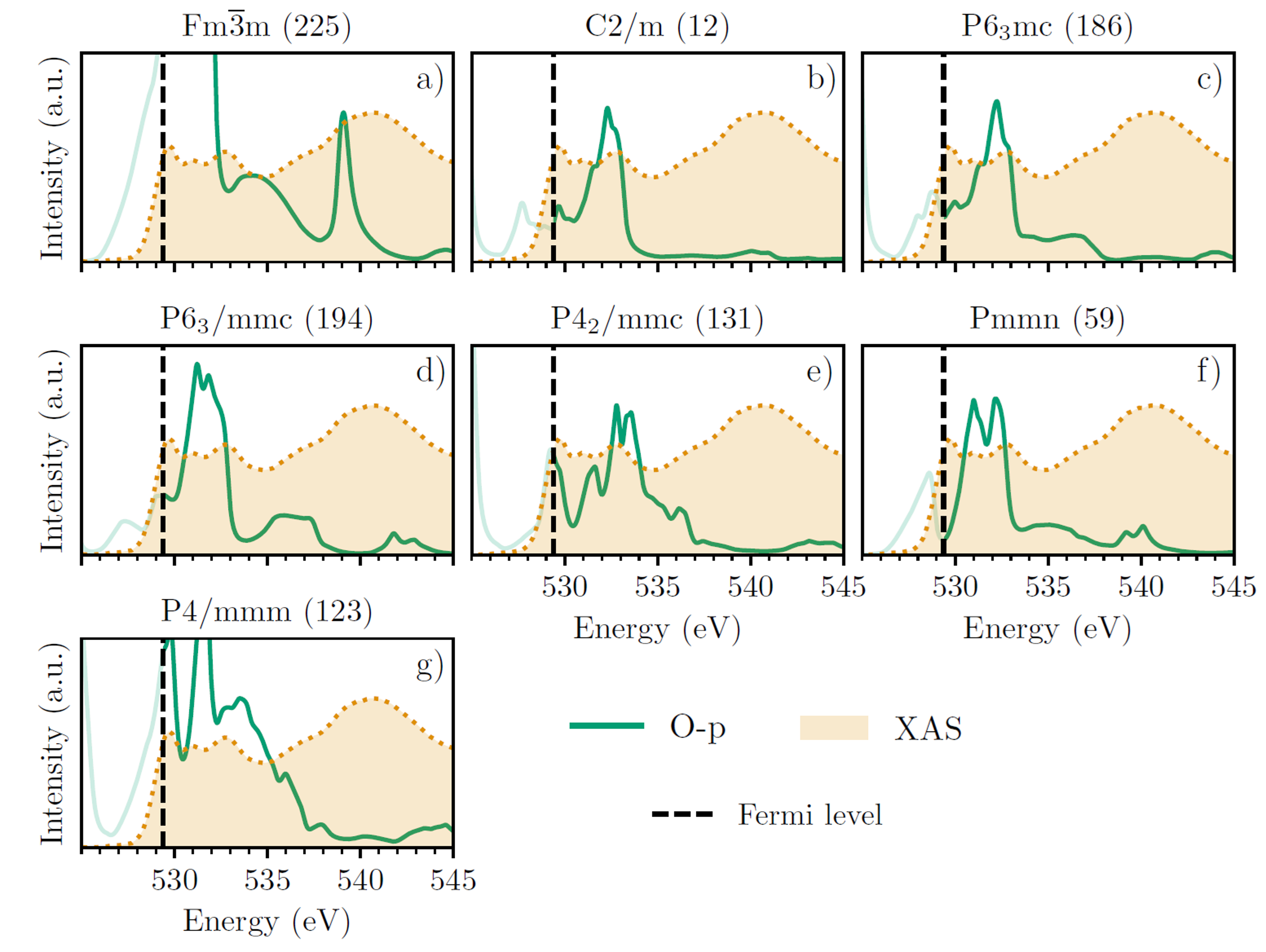}
\caption{Comparison between the calculated unoccupied O $2p$ PDOS for MnO polymorphs (green) and the experimental TEY-XAS spectrum of nominal MnO (red): \textbf{a)}~$Fm\bar{3}m$ (225), \textbf{b)}~$C2/m$ (12), \textbf{c)}~$P6_3mc$ (186), \textbf{d)}~$P6_3/mmc$ (194), \textbf{e)}~$P4_2/mmc$ (131), and \textbf{f)}~$Pmmn$ (59). A Lorentzian broadening of 170~meV is applied to the PDOS curves. All spectra are aligned relative to the calculated Fermi energy by fixing the first conduction peak of the $Fm\bar{3}m$ phase to the experimental onset.}
  \label{fgr:ht_MnO_all}
\end{figure}

The different crystal phases of \ce{MnO} exhibit pronounced variations in the O $p$-PDOS  (Figure~\ref{fgr:ht_MnO_all}). While the ground-state rock-salt phase (Figure~\ref{fgr:ht_MnO_all}a) successfully captures the primary absorption onset of the experimental \ce{MnO} reference spectrum, it reveals distinct and sharp maxima. Moreover, the intensity distribution at higher energy hints at structural contributions from other configurations, such as the cubic (Figure~\ref{fgr:ht_MnO_all}d) and tetragonal (Figure~\ref{fgr:ht_MnO_all}e) phases. In contrast, in the orthorhombic phase (Figure~\ref{fgr:ht_MnO_all}f), the O $p$-contributions produce a strong, broad peak around 531~eV, that aligns well with the experimental spectral onset, accompanied by weaker contributions up to 540~eV that generally track the main experimental features. Finally, the hexagonal $P6_3/mmc$ polymorph (Figure~\ref{fgr:ht_MnO_all}d) accurately captures the intense spectral weight immediately following the absorption threshold between 530 and 533~eV, although it fails to reproduce the broad experimental maximum above 535~eV. Compared with the Co- and Ni-based counterparts, \ce{MnO} exhibits the most complex and structurally sensitive spectral profile. This variation among different polymorphs underscores that no single crystal proxy can fully account for the experimental line shape. Instead, we interpret this structural complexity as a possible ensemble average over multiple, locally distorted coordination environments that emerge during the structural reorganization of the cathode surface layer.

\section{Discussion and Conclusions}
The comparison between experimental XAS data and BSE results across the studied nominal reference oxides demonstrates that while high-level MBPT frameworks provide an accurate description of excitonic effects, they are conventionally applied to idealized structures. Consequently, single-crystal models fail to capture the structural complexity of nominal reference compounds and, by extension, that of real-world battery samples. Our findings demonstrate that the spectral discrepancies observed for the nominal reference oxides are not failures of the many-body electronic framework, but rather distinct spectroscopic signatures of the intrinsic structural heterogeneity present in real materials.

By integrating these high-fidelity BSE benchmarks with a DFT-based high-throughput screening of 38 candidate polymorphs, we resolve this discrepancy. We show that experimental XAS profiles, characterized by broad absorption onsets and altered intensity distributions, are macroscopic manifestations of complex local environments defined by symmetry breaking and lattice strain. While many of these polymorphs are thermodynamically accessible only under extreme bulk conditions, within the context of cathode degradation, they function as effective structural proxies for the metastable, non-equilibrium coordination environments formed during surface reconstruction. Therefore, the experimental spectra are more accurately interpreted as an ensemble average over diverse structural motifs rather than the response of a single idealized lattice.

While the semi-local PBE functional suffers from known, systematic limitations in predicting band gaps of transition-metal oxides, its ability to capture the relative distribution of oxygen $p$-states validates its utility as a rapid descriptor for structural identification. This dual-track approach transforms the limitations of single-phase models into a powerful diagnostic tool: by identifying which local hybridization motifs best match the experimental reference through high-throughput screening, we can prioritize specific, spectroscopically relevant configurations for high-fidelity BSE characterization.

In summary, these results establish that the oxygen K-edge XAS profiles of nominal reference oxides and degraded cathode surfaces should not be interpreted as the signature of a single, stable crystal phase, but rather as an emergent property of local structural heterogeneity. By mapping the O $2p$ fingerprint across a comprehensive library of polymorphs, we provide a robust diagnostic framework that resolves the discrepancies typically encountered in conventional bulk-phase approximations. This strategy is not only powerful for decoding the local chemical and structural evolution of energy materials under \textit{operando} conditions, but also establishes a clear pathway for integration into automated, data-driven modeling workflows for the accelerated characterization of battery materials.

\begin{acknowledgement}
  Advanced Light Source is acknowledged for providing beamtime at the beamline 8.0.1. We thank W. Yang for support during beamtime. Leonhard Reinschl\"ussel and Hubert Gasteiger (Technical University of Munich) are kindly acknowledged for providing lithiated TM oxide samples. 
  This work was funded by the German Federal Ministry for Education and Research - Project No. 03XP0328A and 03XP0328C. C.C. acknowledges additional support from the German Federal Ministry of Education and Research (Professorinnenprogramm III) as well as from the State of Lower Saxony (Professorinnen f\"ur Niedersachsen). The computational resources were provided by the high-performance computing cluster CARL at the University of Oldenburg, funded by the German Research Foundation (Project No. INST 184/157-1 FUGG) and by the Ministry of Science and Culture of the Lower Saxony State, as well as by the North German Supercomputing Alliance, project nip00065.

\end{acknowledgement}


\begin{suppinfo}
Detailed computational parameters, including $k$-grid sampling and muffin-tin radii for all species; a comparison between the BSE and IPA results for representative oxides; schematic overview of the high-throughput workflow used to screen the binary oxide polymorphs and manage the structural uniqueness analysis; a comparative analysis of the O-$p$ PDOS calculated using \texttt{exciting}, \texttt{Quantum ESPRESSO}, and Hubbard-corrected references from the Materials Project.
\end{suppinfo}

\providecommand{\latin}[1]{#1}
\makeatletter
\providecommand{\doi}
  {\begingroup\let\do\@makeother\dospecials
  \catcode`\{=1 \catcode`\}=2 \doi@aux}
\providecommand{\doi@aux}[1]{\endgroup\texttt{#1}}
\makeatother
\providecommand*\mcitethebibliography{\thebibliography}
\csname @ifundefined\endcsname{endmcitethebibliography}  {\let\endmcitethebibliography\endthebibliography}{}

\end{document}